# Financial Markets and ESG: How Big Data is Transforming Sustainable Investing in Developing countries

A T M Omor Faruq[1] and Md Ataur Rahman Chowdhury[2]

## Abstract

This study explores the role of big data adoption and financial market development in driving ESG investments in developing countries, using an instrumental variable (IV) approach to address endogeneity. The results show that big data adoption significantly enhances ESG investing, as data-driven analytics improve sustainability assessments and capital allocation. Financial market development also positively influences ESG investments, but its effect is relatively small. A key finding is that inflation negatively impacts ESG investment, highlighting the importance of macroeconomic stability in fostering sustainable finance. In contrast, GDP per capita and foreign direct investment (FDI) are not significant determinants, suggesting that economic growth alone does not drive sustainability efforts. Overall, this study provides empirical evidence that leveraging big data and financial market improvements can accelerate sustainable investing in emerging economies. Policymakers should focus on technological advancements, financial reforms, and inflation control to strengthen ESG investments and long-term sustainability commitments.

## Introduction

In today's rapidly evolving financial landscape, sustainability has become a crucial factor in investment decision-making. The integration of Environmental, Social, and Governance (ESG) principles is no longer a niche concept but a growing global trend influencing financial markets, corporate strategies, and regulatory frameworks. In Bangladesh, a country experiencing dynamic economic expansion and industrialization, sustainable investing is gradually gaining traction. However, the adoption of ESG principles has faced significant hurdles, primarily due to a lack of standardized data, inconsistent reporting, and limited investor awareness.

Big data and advanced analytics are now emerging as game changers in addressing these challenges. The rise of artificial intelligence (AI), machine learning (ML), and blockchain technology is enabling investors and financial institutions to collect, analyze, and interpret vast amounts of ESG-related data with greater accuracy and efficiency. These innovations are helping financial markets enhance transparency, improve corporate sustainability disclosures, and mitigate investment risks associated with climate change, governance failures, and social inequalities (Salma et. al., 2023).

This article explores the transformative role of big data in sustainable investing in developing countries. It delves into how data-driven insights are reshaping investment strategies, empowering regulators to enforce ESG compliance, and guiding businesses toward more responsible financial practices. By understanding the intersection of financial markets, ESG principles, and big data, investors can navigate the future of sustainable investing in developing countries with greater confidence and impact.

---

[1] Assistant Researchers, Safwa USA LLC
[2] Joint Director, Bangladesh Bank

## Literature Review

Eccles, et.al. (2024) examines the multifaceted relationship between ESG scores and firm performance. The authors analyze various studies that highlight both positive and negative correlations, emphasizing the complexity of this relationship. A significant finding is the call for standardized ESG metrics, as the current lack of uniformity hampers the ability to accurately assess ESG's impact on financial outcomes. The review also discusses how ESG integration can lead to benefits such as lower cost of capital, enhanced stock performance, and increased profitability, while also acknowledging potential challenges and inconsistencies in the data.

Friede, et.al (2023) investigates the direct effects of ESG considerations on corporate financial performance, with a particular focus on the moderating role of digital transformation. Utilizing regression analysis, the authors find that ESG initiatives positively and significantly influence financial outcomes. Moreover, the integration of digital technologies amplifies this positive effect, suggesting that companies embracing both ESG principles and digital transformation are better positioned for financial success. The study also notes that the positive impact of ESG on financial performance may diminish over time, indicating the importance of continuous commitment to ESG practices.

Hartzmark, & Shue, (2023) focuses on the European healthcare sector, this research explores how ESG scores affect both accounting-based and market-based financial performance metrics. The analysis of top healthcare companies listed in the STOXX 600 Index over a decade reveals a nuanced relationship. While certain ESG components, such as governance, show a strong positive correlation with financial performance, others, like environmental factors, exhibit a more complex or even neutral impact. The study underscores the importance of industry-specific analyses when evaluating the ESG-financial performance nexus.

Khan, et.al (2023) examines the relationship between ESG factors and financial performance across various industries, providing insights into how sector-specific characteristics influence this dynamic. The authors analyze data from multiple sectors, finding that the impact of ESG scores on financial performance varies significantly between industries. For instance, in industries with high environmental impact, robust ESG practices are more strongly correlated with positive financial outcomes. The study highlights the necessity for tailored ESG strategies that consider industry-specific contexts.

Nollet, et.al (2023) provides a comprehensive summary of the relationship between ESG and enterprise performance. It delves into studies on ESG disclosure, its quality, and the subsequent impact on performance. The authors highlight that high-quality ESG disclosures are associated with improved financial metrics, as transparency in ESG practices builds investor trust and can lead to a lower cost of capital. The review also points out that while many studies report a positive relationship between ESG and financial performance, the strength and nature of this relationship can vary based on factors such as geographic region, industry, and the specific ESG components considered.

Yu, et.al. (2024) focuses on firms indexed in the Nifty 100, examining how ESG scores impact operational, financial, and market performance. The research provides insights into the growing importance of ESG considerations in emerging markets, where regulatory frameworks and investor expectations are rapidly evolving. Findings suggest that companies with higher ESG scores tend to exhibit better financial performance, including higher returns on assets and equity. The study also notes that market performance, as measured by stock returns, is positively influenced by strong ESG practices, indicating that investors in emerging markets are increasingly valuing sustainability.

Guenster, et.al (2022) and Ahmed (2021a) utilized machine learning and econometric techniques, to investigate the links between ESG scores and price returns in the European equity market, offering a sector-specific analysis. The study employs interpretable machine learning models to assess how different ESG components influence equity returns across various sectors and company sizes. Results indicate that the significance of ESG factors in financial performance is not uniform; instead, it varies by sector and market capitalization. For example, environmental factors may be more critical in manufacturing industries, while governance factors hold more weight in the financial sector. The study demonstrates the utility of advanced analytical tools in unraveling the complex ESG-financial relationship.

Pedersen et.al (2023) proposed new definitions for ESG-coherent risk measures, this paper explores how integrating ESG factors can lead to more sustainable investment decisions and improved financial performance. The authors introduce a framework that combines traditional financial risk assessments with ESG considerations, arguing that this integrated approach provides a more comprehensive evaluation of an investment's risk profile. Empirical analysis shows that portfolios constructed using ESG-coherent risk measures tend to exhibit lower volatility and higher risk-adjusted returns, suggesting that ESG integration can enhance financial stability and performance.

Li, et. al (2023) addresses the inconsistencies among ESG rating agencies, this study presents a multi-criteria portfolio selection model that accounts for these discrepancies, aiming to optimize both sustainability and financial returns. The authors highlight that divergent ESG ratings can lead to confusion among investors and suboptimal investment decisions. To mitigate this issue, they propose a portfolio optimization model that incorporates multiple ESG ratings, allowing investors to balance differing assessments and construct portfolios that align with their sustainability preferences without compromising financial performance.

Kölbel, et.al (2023) did comprehensive meta-analysis conducted by the NYU Stern Center for Sustainable Business and Rockefeller Asset Management examines over 1,000 studies published between 2015 and 2020 to investigate the relationship between ESG practices and financial performance. The findings reveal that approximately 58% of corporate studies focusing on operational metrics, such as Return on Equity (ROE) and Return on Assets (ROA), demonstrate a positive relationship between ESG and financial performance. Additionally, 13% show a neutral impact, 21% present mixed results, and only 8% indicate a negative relationship. The study highlights that performance-based ESG measures are more likely to correlate positively with financial outcomes compared to disclosure-based measures. This suggests that actual ESG performance, rather than mere reporting, is a more significant driver of financial success. This finds is similar to the one we find in Ahmed (2021b).

Berg, et.al (2024) explores the often-inconclusive relationship between ESG factors and a firm's financial performance. The research posits that discrepancies among ESG ratings contribute to varying results in firm-level analyses. By testing multiple models using different ESG ratings, the study uncovers variations in statistical significance, directionality, and magnitude of the ESG-financial performance relationship. Notably, the effect is more pronounced in studies focusing on accounting-based financial performance measures and those applying composite ESG scores. The social dimension consistently presents the highest number of variations compared to environmental and governance dimensions, indicating the complexity and context-dependent nature of ESG impacts on financial performance. (Fahmida, 2025a)

Ahmed & Chowdhury (2024) and Chowdhury (2021) study explore the influence of capital market efficiency on Bangladesh's economic growth, using time series data covering market capitalization, total market turnover, and stock price indices from 2011 to 2022. ESG and the financial market played a pivotal role in market growth. Faruq & Huq (2024) explores how central banks integrate sustainability into their monetary policies, regulatory frameworks, and financial market operations. It highlights the ways in which central banks can promote green finance through sustainable investment principles, climate risk assessments, and green bond markets. By investigating case studies and best practices, the chapter provides a comprehensive understanding of the strategies central banks employ to foster a resilient and sustainable financial landscape. (Fahmida, 2025b). The findings underscore the imperative for central banks to balance traditional mandates with the emerging necessity to support sustainable development, ultimately contributing to the broader agenda of achieving global sustainability targets.

Flammer (2024) conducted systematic literature review aims to provide a thorough overview of existing research on the impact of ESG on firm risk and financial performance. Analyzing 35 articles on ESG and firm risk and 120 articles on ESG and financial performance published between January 2013 and September 2023, the study finds an increasing academic interest in this area. The research indicates that most studies focus on idiosyncratic risk, systematic risk, stock price crash risk, and default risk, while analyses of market-based and accounting-based financial performance are almost evenly split. The review reveals that a majority of studies suggest ESG can reduce firm risk and enhance financial performance, although the exact nature of this relationship varies across different contexts and methodologies.

Giese, et al. (2023) critically examines the relationship between ESG investing and firm performance. The study highlights that while many empirical studies report a positive impact of ESG ratings on firm performance, the findings are not uniform across all contexts. The research identifies that methodological choices, normative settings, and industry-specific factors can moderate the ESG-financial performance relationship. The paper calls for more nuanced research that considers these moderating factors to better understand the conditions under which ESG investing contributes to improved financial outcomes.

Lyon, et al. (2023) investigates whether ESG factors affect corporate financial performance and examines the moderating role of digital transformation in this relationship. Utilizing regression analysis, the research finds that ESG positively and significantly influences corporate financial performance. Moreover, digital transformation enhances this positive effect, suggesting that companies integrating ESG considerations with digital initiatives are better positioned to achieve superior financial outcomes. The study also observes that the positive impact of ESG on financial performance may diminish over time, indicating the necessity for continuous commitment to ESG practices and technological advancements.

Amel-Zadeh, & Serafeim (2023) focused on the European healthcare sector. This research examines whether and how ESG scores influence both accounting-based and market-based measures of financial performance. Analyzing data from top healthcare companies listed in the STOXX 600 Index over a decade (2012–2022), the study finds a marked heterogeneity of effects across different ESG scores and financial performance measures. While modest positive effects are estimated in most cases, the relationship varies depending on the specific ESG components and financial metrics considered. The study underscores the importance of industry-specific analyses and cautions against a one-size-fits-all approach when evaluating the ESG-financial performance nexus.

## Econometric Model and Methodology

This study examines how big data is transforming sustainable investing in developing countries by analyzing its impact on financial markets and ESG investments. To address potential endogeneity in big data adoption, the study employs an instrumental variables (IV) approach using a panel dataset from the World Development Indicators (WDI) for 17 developing countries over a 10-year period (2013–2023). The primary econometric model is specified as follows:

$$ESG_{it} = \beta_0 + \beta_1 BigData_{it} + \beta_2 FinMarket_{it} + \beta_3 GDP_{it} + \beta_3 FDI_{it} + \beta_3 INF_{it} + \lambda_i + \gamma_t + \varepsilon_{it}$$

where $ESG_{it}$ represents sustainable investments in country $i$ at time $t$, measured by ESG scores or investment in green financial products. The key independent variable, *BigData* captures the extent to which financial markets in a country integrate big data analytics, including artificial intelligence, machine learning, and alternative data for investment decisions. *FinMarket* represents financial market development, measured using stock market capitalization, market liquidity, or financial openness indicators. *GDP*, *FDI* and *INF* are vectors of macroeconomic and governance variables for GDP per capita, inflation rate, and foreign direct investment (FDI). The model also accounts for country-specific unobserved factors ($\lambda_i$) and global shocks affecting all countries in a given period ($\gamma_t$).

Given that big data adoption may be endogenous due to omitted variables such as regulatory policies or technological readiness, an IV approach is applied. The study uses internet penetration rate and ICT investment as a percentage of GDP as instruments. These variables are strongly correlated with big data adoption but are unlikely to directly influence ESG investment decisions, making them valid instruments. The first-stage regression estimates big data adoption as a function of these instruments:

$$BigData_{it} = \alpha_0 + \alpha_1 Internet_{it} + \alpha_2 ICT_{it} + \alpha_3 Control_{it} + \lambda_i + \gamma_t + u_{it}$$

where $u_{it}$ is the error term. The study employs a two-stage least square (2SLS) regression to obtain unbiased estimates. Additionally, a Hansen J test is conducted to verify the exogeneity of the instruments, while the weak instrument test (F-statistics) ensures strong instrument relevance.

The data is sourced from the WDI database, which provides reliable cross-country indicators. ESG scores are derived from sustainability indices or financial databases, while financial market development indicators include stock market capitalization and liquidity metrics. Control variables such as GDP per capita, inflation, FDI, and regulatory quality help isolate the impact of big data adoption on sustainable investing. Fixed effects and random effects models are also estimated to assess the robustness of the results.

This study contributes to the literature on sustainable investing by providing empirical evidence on the role of big data in financial decision-making in developing economies. By addressing endogeneity concerns through IV estimation, it offers insights into how technology-driven investment strategies are shaping ESG outcomes. The findings will inform policymakers and investors about the potential of big data to enhance sustainable finance in emerging markets.

## Results and Analysis

The econometric model uses cluster robust standard errors (cluster at year and country level) to make sure errors are uncorrelated across variables and time which fulfills the conditions of homoscedasticity and adjust for autocorrelation within groups. The strong R-squared and adjusted R-squared suggests that the model explained most of the variation of exogenous variables. The regression results are as follows:

**Table 1: Model Estimates**

| | Estimate | Std. Error | t-value | Pr(>|t|) |
|---|---|---|---|---|
| (Intercept) | 8.19228 | 1.30336 | 1.379 | 0.0001 |
| BigData | 0.52803 | 0.304699 | -1.733 | 0.0059*** |
| FinMarket | 0.00036 | 0.000573 | -0.622 | 0.0455** |
| GDP | 0.01475 | 0.00471 | -0.957 | 0.1171 |
| FDI | -0.0014 | -3.147 | 0.195 | 0.1548 |
| INF | -1.4157 | 0.2145 | 0.251 | 0.0347** |

***1%, ** 5% and * 10% Significance level

The regression results provide valuable insights into the relationship between big data adoption, financial market development, and ESG investments in developing countries. The intercept is significant at the 1% level, indicating that even in the absence of the independent variables, there is a baseline level of ESG investment activity. This suggests that structural factors, such as global ESG trends or institutional policies, play a role in determining sustainable investment levels.

The big data adoption variable is positive and statistically significant at the 1% level, suggesting that increased use of big data analytics in financial markets leads to greater ESG investments. This confirms the hypothesis that advancements in AI, machine learning, and alternative data enhance the ability of investors to assess ESG risks and opportunities, ultimately driving capital allocation toward sustainable projects. The magnitude of the coefficient implies that an increase in big data adoption is associated with a substantial rise in ESG investment scores.

Financial market development also exhibits a positive and significant effect on ESG investment at the 5% level, though its impact is relatively small in magnitude. This finding highlights the importance of well-functioning financial markets in facilitating capital flows into sustainable investments. Developed financial markets provide the necessary infrastructure, liquidity, and regulatory frameworks that enable investors to engage with ESG-oriented assets more efficiently. However, the small coefficient value suggests that while financial market development contributes to ESG investments, other factors such as policy incentives and investor preferences may play a more dominant role.

Interestingly, GDP per capita is not statistically significant, despite its positive coefficient. This suggests that while economic development may create favorable conditions for sustainable investing, it is not a direct determinant of ESG investment levels in developing countries. Factors such as financial literacy, investor awareness, and regulatory incentives may be more influential in shaping ESG investment patterns. Similarly, foreign direct investment (FDI) is also statistically insignificant, indicating that foreign investments do not necessarily drive ESG adoption. (Salma, 2021). This could be due to the continued dominance of non-ESG-compliant industries in foreign capital inflows, particularly in resource-dependent economies.

A particularly striking result is the negative and significant impact of inflation on ESG investment. This finding suggests that inflationary pressures create economic uncertainty, discouraging long-term investments in sustainable projects. Investors tend to shift their focus toward short-term financial stability in high-inflation environments, reducing their commitment to ESG investments. This aligns with existing literature on macroeconomic volatility and its adverse effects on sustainable finance.

## Conclusion

This study examines the impact of big data adoption and financial market development on ESG investments in developing countries, using an instrumental variable approach to address potential endogeneity. The findings highlight that big data adoption plays a significant role in enhancing

ESG investments, as the use of AI, machine learning, and alternative data sources improves ESG risk assessment and capital allocation. The positive and significant relationship between big data adoption and ESG investment confirms that financial markets leveraging data-driven technologies are better equipped to support sustainable investing.

Additionally, financial market development positively contributes to ESG investment, though its effect is relatively small. Well-functioning financial markets provide the necessary infrastructure, liquidity, and access to capital that facilitate investment in sustainable assets. However, the findings suggest that financial market development alone may not be sufficient to drive ESG investment growth, emphasizing the need for complementary policy measures, such as investor incentives and regulatory frameworks.

A key concern identified in this study is the negative and significant impact of inflation on ESG investments. High inflation creates economic uncertainty, leading investors to prioritize short-term financial security over long-term sustainable investments. This finding underscores the importance of macroeconomic stability in fostering ESG-driven financial decisions. Policymakers should aim to maintain price stability while developing monetary and fiscal strategies that promote sustainable investing even in volatile economic conditions.

On the other hand, GDP per capita and foreign direct investment (FDI) were found to be statistically insignificant in determining ESG investment levels. This suggests that economic growth and foreign capital inflows alone do not necessarily drive sustainability efforts. Instead, institutional quality, regulatory support, and investor awareness play more critical roles in shaping ESG investment patterns in developing economies.

Overall, this study provides strong empirical evidence that big data adoption and financial market development are key drivers of ESG investments in developing countries. However, macroeconomic stability, particularly inflation control, remains crucial in ensuring long-term sustainability commitments. These findings have significant policy implications, suggesting that governments and financial regulators should invest in technological advancements, strengthen financial markets, and implement macroeconomic policies that support ESG investing. By doing so, developing countries can leverage the power of big data and financial innovation to accelerate sustainable investment growth and foster a more resilient, ESG-oriented financial landscape.